\documentstyle[12pt,epsfig]{article}
\begin{document}
\large
\begin{center}
{\Large\bf Leading Effects in the Spectra of  $\Lambda_c$ and 
$\bar{\Lambda}_c$
Produced in $\Sigma^-p$, $pp$ and $\pi^-p$ Interactions.}

\vspace{4mm}

 O.I.Piskounova\\

 P.N.Lebedev Physical Institute of Russian Academy of Science,

Moscow \\

 Russia \\

\end{center}

\begin{abstract}

The spectra of leading and nonleading charmed baryons ($\Lambda_c$ 
and $\bar{\Lambda}_c$) as well as the asymmetries between these spectra 
measured in $\Sigma^-A$, $\pi^-A$ and pA collisions 
at $p_L= 600 GeV/c$ in the E781 experiment are simultaneously 
described within 
the framework of Quark-Gluon String Model (QGSM). It is shown that the 
charmed baryon spectra can be fitted by QGSM curves calculated 
with the parameter of diquark fragmentation, $a_f^{\Lambda_c}$=0,006. 
It was obtained in this experiment that the
asymmetry between the spectra of $\Lambda_c$ and $\bar{\Lambda}_c$ 
in $\pi^-A$ collisions is of nonzero value. It might be 
described in our model only assuming that the string junction is 
transfered from target proton into the kinematical region 
of pion projectile fragmentation. 

\end{abstract}

\newpage

\section{Introduction}

The data of E781 experiment \cite{selex}(FNAL) on
spectra of charmed baryons as well as the asymmetries between 
$\Lambda_c$ and $\bar{\Lambda}_c$ in $\Sigma^-A$, $\pi^-A$
and pA interactions at $p_L= 600 GeV/c$ have recently amplified 
the results of WA89 experiment \cite{wa89} (CERN)
at $p_L= 340 GeV/c$  and E791 experiment \cite{e791}(FNAL) at 
$p_L= 500 GeV/c$. The data of these experiments on charmed 
meson spectra and asymmetries have been already considered in 
the recent paper \cite{Dasym} from the point of view of 
Quark Gluon String Model (QGSM) in order to understand the 
influence of quark composition of beam particle on the shape 
of production spectra of heavy flavored particles. 

The difference in $x$ spectra ($x=x_F=2p_{II}/\sqrt{s}$) of 
leading and nonleading particles  has been explained successfully 
by several theoretical models as an effect of interplay between 
the quark contents of the projectile and of the produced hadron.

Most advanced QCD models \cite{ic,r2c} have to take into account 
so called "intrinsic charm" (IC) in order to decribe the high value of 
asymmetry between
the $x$-spectra of charmed particles and antiparticles in the 
fragmentation region, $x\rightarrow$1. In QGSM \cite{qgsm} and 
other models \cite{likhoded} with elaborated concepts of fragmentation 
functions (FF),
there is no necessity of such assumption, because the specifically 
written FF give the asymmetries rising with $x$. Some amount of
IC can only suppress the asymmetry, as it was shown in
previous calculations \cite{Dasym,arakelyan}. 

It should be noticed that there is a large difference between 
leading effects in charmed meson spectra and those effects in charmed baryon 
spectra. Leading $\Lambda_c$ baryon in proton-proton 
interaction might be produced 
by the "leading" fragmentation of proton $ud$-diquark bringing the large fraction of
proton momentum that gives an important enchancement of $\Lambda_c$ spectra 
over the spectra of $\bar{\Lambda}_c$ . We are suggesting here to consider 
the spectra in full $x$ region: $-1<x<1$, so that the left side of 
plots always corresponds  to target proton fragmentation. For example, in the case of 
hyperon-proton interactions the $\Lambda_c$  spectra will have two different "wings": 
left one at the negative $x$'s shows the high 
asymmetry towards $\bar{\Lambda}_c$ spectra due to the leading 
fragmentation of target proton, the other one in the positive $x$ 
region should have a bit lower asymmetry because the fragmentation of hyperon diquarks 
is supposed to have not that strongly leading character \cite{lambda_c} as proton 
$ud$-diquark fragmentation.

The asymmetry between $\Lambda_c$ and $\bar{\Lambda}_c$ spectra 
in pion-proton interaction should be equal zero in the 
region of pion fragmentation because pion can have the valent quark 
(antiquark) in common as with $\Lambda_c$ as with $\bar{\Lambda}_c$, so
both spectra will provide a leading character and be equal.

\section{Valence Quark Distributions in QGSM.}

The inclusive production cross section of hadrons of type H is written as
a sum over n-Pomeron cylinder diagrams:

\begin{equation}
f_{1}=x \frac{d \sigma^{H}}{dx}(s,x)= \int E \frac{d^{3}\sigma^{H}}
{d^{3}p}d^{2}p_{\bot}=\sum_{n=0}^{\infty}\sigma_{n}(s) \varphi_{n}^{H}(s,x).
\end{equation}

Here, the function  $\varphi_{n}^{H}(s,x)$ is a particle distribution in the
configuration of n cut cylinders and $\sigma_{n}$ is the probability of
this process. The cross sections $\sigma_{n}$ depend  on 
parameter of the supercritical Pomeron $\Delta_P$, which is equal in our model
to 0,12 \cite{qgsm}. 

The distribution functions of $\Lambda_c$ in case of $\pi^-p$ collisions
 are given by:

\begin{eqnarray}
\varphi_{n}^{\Lambda_c}(s,x)=a_{0}^{\bar{\Lambda}_c}
[F_{q}^{(n)}(x_{+})F_{qq}^{(n)}(x_{-})+
F_{qq}^{(n)}(x_{+})F_{q}^{(n)}(x_{-})+ \\
2(n-1)F_{q_{sea}}^{(n)}(x_{+})F_{\bar{q}_{sea}}^{(n)}(x_{-})]+
a_f^{\Lambda_c}F_{1qq}^{(n)}(x_{-}),\nonumber
\end{eqnarray}
where $a_0^{\bar{\Lambda}_c}$ is the central (vacuum) density parameter 
of charmed baryon production and $a_f^{\Lambda_c}$ is the fragmentation 
parameter of proton target diquark.
In the case of $\Lambda_c$ production in proton fragmentation the diquark 
fragmentation plays an important role, this diquark part of distribution should be 
written separately. So the distribution for $pp$ collision will include 
two diquark parts, as for positive $x$ as for negative:

\begin{eqnarray}
\varphi _{n}^{\Lambda_c}(s,x)=a_f^{\Lambda_c}F_{1qq}^{(n)}(x_{+})+
a_f^{\Lambda_c}F_{1qq}^{(n)}(x_{-})+
a_{0}^{\bar{\Lambda}_c}[F_{q}^{(n)}(x_{+})F_{0qq}^{(n)}(x_{-})+ \\
F_{0qq}^{(n)}(x_{+})F_{q}^{(n)}(x_{-})+ 
2(n-1)F_{q_{sea}}^{(n)}(x_{+})F_{\bar{q}_{sea}}^{(n)}(x_{-})],\nonumber 
\end{eqnarray}
where $F_{1qq}^{(n)}(x_{+})$ is the distribution at the leading fragmentation
of diquarks, while $F_{0qq}^{(n)}(x_{+})$ is the ordinary part of fragmentation
written with the central density parameter $a_{0}^{\bar{\Lambda}_c}$.

The $\Lambda_c$ distribution functions in case of 
$\Sigma^-p$ collisions includes also the additional diquark parts:

\begin{eqnarray}
\varphi_{n}^{\Lambda_c}(s,x)=a_f^{\Lambda_c}F_{1qq}^{(n)}(x_{+})+
a_f^{\Lambda_c}F_{1qq}^{(n)}(x_{-})+ \\
a_{0}^{\bar{\Lambda}_c}[F_{q}^{(n)}(x_{+})F_{qq}^{(n)}(x_{-})+
F_{qq}^{(n)}(x_{+})F_{q}^{(n)}(x_{-})
+2(n-1)F_{q_{sea}}^{(n)}(x_+)F_{\bar{q}_{sea}}^{(n)}(x_{-})],\nonumber
\end{eqnarray}
where $a_0^{\bar{\Lambda}_c}$ and $a_f^{\Lambda_c}$ are the same density 
parameters as in eqs.(2) and (3).

The particle distribution on each side of chain can be built on the
account of quark contents of beam particle ($x_+~=~(x+\sqrt{x^2+x_{\bot}^2})/2$,
$x_{\bot}=2\sqrt{m_{\Lambda_c}^2+\bar{p_{\bot}^2}}/\sqrt{s}$) and  of target particle ($x_-~=~(x-\sqrt{x^2+
x_{\bot}^2})/2$). They are in a case of $\Sigma^-p$ collisions: 

\begin{eqnarray}
 F_{q}^{(n)}(x_{+})&=&\frac{1}{3}F_{s}^{(n)}(x_{+})+\frac{2}{3}F_{d}^{(n)}(x_+),\nonumber \\
 F_{qq}^{(n)}(x_{+})&=&\frac{1}{3}F_{dd}^{(n)}(x^{+})+\frac{2}{3}F_{ds}^{(n)}(x_{+}), \\
 F_{q}^{(n)}(x_{-})&=&\frac{1}{3}F_{d}^{(n)}(x_{-})+\frac{2}{3}F_{u}^{(n)}(x_{-}),\nonumber \\
 F_{qq}^{(n)}(x_{-})&=&\frac{1}{3}F_{uu}^{(n)}(x_{-})+\frac{2}{3}F_{ud}^{(n)}(x_{-}).\nonumber
\end{eqnarray}

Each $F_{i}(x_{\pm})$ ($i=s,u,d,ud,dd,ds$...) is constructed as the convolution:
\begin{equation}
 F_{i}(x_{\pm})=\int_{x_{\pm}}^{1} f_{\Sigma^-}^{i}(x_{1})\frac{x_{\pm}}{x_{1}}
 {\cal D}_{i}^{H}(\frac{x_{\pm}}{x_{1}})dx_{1},
\end{equation}
where $f^{i}(x_{1})$ is a structure function of i-th quark ( diquark or antiquark) 
which has a fraction of energy $x_{1}$ in the interacting hadron  and 
${\cal D}_{i}^{H}(z)$
is a fragmentation function of this quark into the considered type
of produced hadrons H.

The structure functions of quarks in interacting proton, hyperon, or pion  
beams have already been described
in the previous papers \cite{oldcharm,sigmabeam,pionbeam}. In the case of 
hyperon beam they depend
on the parameter of the Regge trajectory of $\varphi$-mesons ($s\bar{s}$)
because of s-quark contained in $\Sigma^-$
($\alpha_{\varphi}$(0)=0).

\section{Diquark Fragmentation Function and String Junction Transfer}

The fragmentation functions of diquark and quark chains into charmed baryons 
or antibaryons are based on the rules written in \cite{kaidalov}.

The $ud$- and $dd$-diquark fragmentation function includes the constant $a_f^{\Lambda_c}$ 
which could be interpreted as "leading" parameter, but the value of $a_f^{\Lambda_c}$ 
is fixed due to the baryon number sum rule and should be approximately equal to the value taken for 
$\Lambda_c$ spectra in our previous calculations \cite{oldcharm}:

\begin{equation}
{\cal D}_{dd}^{\Lambda_c}(z)=\frac{a_f^{\Lambda_c}}{a_0^{\bar{\Lambda}_c}z}
z^{2\alpha_R(0)-2\alpha_N(0)}
(1-z)^{-\alpha_{\psi}(0)+\lambda+2(1-\alpha_R(0))},
\end{equation}
where the term $z^{2\alpha_R(0)-2\alpha_N(0)}$ means the probability for
initial diquark to have z close to 0; the intercepts of Regge trajectories,
$\alpha_R(0)$, $\alpha_N(0)$ and $\alpha_{\psi}(0)$ are taken of the same 
values as in \cite{oldcharm}, 0.5, -0.5 and -2.0 correspondingly. The
$\lambda$ parameter is an remnant of transverse momenta dependence, it is equal to 0,5
here ( for more information see the early publications \cite{qgsm,oldcharm}).

It is important here to keep in mind the possibility to create the $\Lambda_c$
baryon only on the  base of  string junction taken from interacting proton or 
$\Sigma^-$. The string junction  brings 
the positive baryon number in baryons and the negative one in antibaryons. In the
proton and hyperon reactions we have diquarks, so only positive baryon number should
be transfered. 
The fragmentation function of string junction that can be transfered to region $z>0$
is of the similar form as diquark FF written above, eq. (7) :
  
\begin{equation}
{\cal D}_{SJ}^{\Lambda_c}(z)=\frac{a_f^{\Lambda_c}}{a_0^{\bar{\Lambda}_c}z}
z^{1-\alpha_{SJ}(0)}(1-z)^{-\alpha_{\psi}(0)+\lambda+2(1-\alpha_R(0))},
\end{equation}
where $\alpha_{SJ}(0)$ is the intercept of string junction Regge trajectory. 
We are not discussing here the two possible values 
of string junction intercept: 0,5 \cite{strjunction} and 1,0 \cite{kopeliovich} 
just taking it  equal to 0,5.
This choice of the intercept is a reson of the target proton string junction going  
easier into 
the region of opposite z than 
the diquark, as it is seen from the comparison of $z\rightarrow0$ asymptotics 
in the last formulas. 
It will become significant when we study the baryon spectra in pion interactions. The full list of 
fragmentation functions of diquarks and of string junction into charmed 
baryons is presented in Appendix.

The main difference between the concepts of asymmetry for D-meson production 
\cite{Dasym} and for
$\Lambda_c$ production is the difference between the forms of leading 
fragmentation functions. The parameter $a_1$, which was taken for the leading 
fragmentation of valence quark into D-mesons ( see \cite{Dasym}), is the ratio of leading D-meson density
in fragmentation region, $z \rightarrow 1$, to the density
in the central region, $z \rightarrow 0$. The $a_f^{\Lambda_c}$ parameter is an absolute 
fraction of the energy of diquark that is brought by produced $\Lambda_c$. But 
both parameters reflect actually  the same idea of high density of leading 
hadrons near the fragmentation region ($z\rightarrow1$) of those quarks (diquarks) of
beam particle which can go into content of this leading hadron. 
This phenomenon was also named a "beam drag" effect in some publications.

\section{Sea Quark Fragmentation Functions}

The main peculiarity of QGSM is the multiple pomeron exchanges \cite{qgsm} those are taken 
into account at the calculations of the spectra of multipartile production, eg.(1). 
In this case the 2(n-1) quark-antiquark chains are connected to
paired sea quark-antiquarks of the beam and target particles.

The structure functions of sea
quark pairs can be written in the same way 
as the valence quark distributions. The  structure function of $d$-quark
in hyperon, for example, is the following:

\begin{eqnarray}
f_{\Sigma^{-}}^{d}(x_{1})&=&C_{d,\bar{d}}^{(n)}x_{1}^{-\alpha_{R}(0)}(1-x_{1})^
{\alpha_{R}(0)-2\alpha_{N}(0)+(\alpha_{R}(0)-\alpha_{\varphi}(0))+
n-1+2(1-\alpha_{R}(0))}.
\end{eqnarray}

Here, sea quarks and antiquarks have an additional power term
$2(1-\alpha_{R}(0))$ corresponding to the quark distribution 
of two pomeron diagram that is including one sea quark pair.   

The fragmentation functions of light $u,d$ sea quark fragmentation into $\Lambda_c$ as well as 
$\bar{u},\bar{d}$ quark into $\bar{\Lambda}_c$ are easily built from valence
quark fragmentation functions. They are also written in the 
Appendix.  

\section{Spectra and Asymmetry of $\Lambda_c$/$\bar{\Lambda}_c$ in 
$\pi^-p$ collisions}

The asymmetry between the spectra of $\Lambda_c$ and $\bar{\Lambda}_c$ measured in
$\pi^-A$ collisions at $p_L$= 600 GeV/c \cite{selex} is shown in Fig.1 a). The nonzero 
asymmetry in the region of pion fragmentation is described on the base of baryon string 
junction transfer from the proton fragmentation region (see section 3).  

The asymmetry is defined as:
\begin{equation}
A(x)=\frac{dN^{\Lambda_c}/dx-dN^{\bar{\Lambda}_c}
/dx}{dN^{\Lambda_c}/dx+dN^{\bar{\Lambda}_c}/dx},
\end{equation}
Here $dN^{\Lambda_c}/dx$ and $dN^{\bar{\Lambda}_c}/dx$ 
are the event distributions measured in the experiment \cite{selex}. 

The invariant distributions $xdN/dx$ of charmed baryons and antibaryons obtained in pion interactions
in E781 experiment are shown in Fig.1 b) with the QGSM curves calculated for
pion fragmentation (the side of positive $x$) and for proton fragmentation (the side of
negative $x$). The ratio between the values of $xdN/dx$($p\rightarrow\Lambda_c,\bar{\Lambda}_c$) and
$xdN/dx$($\pi^-\rightarrow\Lambda_c,\bar{\Lambda}_c$)depends on the ratio of cross 
sections of these two reactions. The absolute values of cross sections are not 
measured in the present experiment, so the left side of experimental plot in 
Fig.1 b) can be shifted towards the right side by the arbitrary factor, and we did 
it here in order to make a better description.

\section{The Spectra and Asymmetry of $\Lambda_c$/$\bar{\Lambda}_c$ in 
$\Sigma^-p$ collisions}

The asymmetry between the spectra of $\Lambda_c$ and $\bar{\Lambda}_c$ measured in
$\Sigma^-A$ collisions at $p_L$= 600 GeV/c is shown in Fig.2 a). Asymmetry is high
in both sides of graph because the diquark fragmentation takes place for the 
beam and the target particles. 

The invariant distributions $xdN/dx$ of charmed baryons and antibaryons obtained in hyperon interactions
in E781 experiment are shown in Fig.2 b) with the QGSM curves calculated as for
hyperon fragmentation (the side of positive $x$) as for proton fragmentation 
(the side of negative $x$). The ratio between the values of $xdN/dx$($p\rightarrow\Lambda_c,\bar{\Lambda}_c$) and
$xdN/dx$($\Sigma^-\rightarrow\Lambda_c,\bar{\Lambda}_c$)depends on the ratio of cross 
sections of this two reactions. The left side of experimental plot is shifted 
toward the right side by the arbitrary factor for the better description as we did 
in the case of pion reaction. 

The complete calculations carried out with the fragmentation 
function written for $\Lambda_c$ and $\bar{\Lambda}_c$ production
give the good description of data with the value of parameter 
$a_f^{\Lambda_c}$=0,006. 

\section{Conclusions}

In this paper we have examined the data on charm baryon production in proton,
pion and hyperon beam interactions with the fixed target at $p_L= 
600 GeV/c$ in the E781 experiment. The following new ideas about $\Lambda_c$ 
and $\bar{\Lambda}_c$ spectra 
and asymmetries are to be mentioned here as the outcome of the QGSM study:

	a)the features of baryon charge transfer by the string junction 
          of the target proton  are disclosed in the nonzero
	  baryon/antibaryon asymmetry in the pion beam fragmentation
	  region although we did not intend here to distinguish  
          between two values of $\alpha_{SJ}(0)$;

	b)$\Lambda_c$ and $\bar{\Lambda}_c$ spectra in the proton 
          and hyperon beam interactions can be described 
	  with the same leading fragmentation parameter, 
          $a_f^{\Lambda_c}$=0,006;

	c)the asymmetry is not a proper quantity to study the bechavior of
	  baryon spectra in the region of $x$ close to 1; though
          the baryon/antibaryon asymmetry for $\pi-p$ reaction  
          shows the good agreement with QGSM curves, the spectra of
          charmed baryons require more detailed description in pion 
          fragmentation region;

	d)there is no necessity to  involve the intrinsic charm  into the 
          calculations of charmed baryon spectra at the up-to-date level of
          experimental data.

Author would like to express her grattitude to Prof.A.B.Kaidalov, 
Dr. M.Iori, Dr. F.Garcia and Dr.S.Baranov for the numerous discussions. 
This work was supported by NATO grant PST.CLG.977021.

\section {Appendix}
\setcounter{equation}{0}
\renewcommand{\theequation}%
{A.\arabic{equation}}
The concept of quark chain fragmentation function has been manifested in 
\cite{qgsm,kaidalov}. The production of $\Lambda_c$ as well as $\bar{\Lambda}_c$
(${\cal D}_{0ud}^{\bar{\Lambda}_c}(z)$ etc.) takes place in the central 
region ($z$=0) of quark-antiquark chain with 
the constant density parameter $a_0^{\bar{\Lambda}_c}$= 4,0$10^{-4}$. The fragmentation 
functions of projectile diquarks into $\Lambda_c$ (${\cal D}_{1ud}^
{\Lambda_c}(z)$ and the similar) 
require the specific parameter $a_f^{\Lambda_c}$=0,006. Departing from these statements the 
full set of FF that is necessary
for the calculation of spectra of $\Lambda_c$ and $\bar{\Lambda}_c$ is written 
as following:

\begin{eqnarray}
&&{\cal D}_{u}^{\Lambda_c}(z)={\cal D}_{d}^{\Lambda_c}(z)=
\frac{a_0^{\bar{\Lambda}_c}}{z}(1-z)^{\alpha_{R}(0)-2\alpha_{N}(0)+\lambda+
\alpha_R(0)-\alpha_{\psi}(0)}, \nonumber\\
&&{\cal D}_{\bar{u}}^{\Lambda_c}(z)={\cal D}_{u}^
{\bar{\Lambda}_c}(z)=\frac{a_0^{\bar{\Lambda}_c}}{z}(1-z)^{\alpha_{R}(0)-2\alpha_{N}(0)+
\lambda+\alpha_R(0)-\alpha_{\psi}(0)+2(1-\alpha_R(0))}, \nonumber\\
&&{\cal D}_{s}^{\Lambda_c}(z)=\frac{a_0^{\bar{\Lambda}_c}}{z}
(1-z)^{\alpha_{R}(0)-2\alpha_{N}(0)+\lambda+
\alpha_R(0)-\alpha_{\psi}(0)+2(1-\alpha_R(0))+
\alpha_R(0)-\alpha_{\phi}(0)}, \nonumber\\
&&{\cal D}_{1ud}^{\Lambda_c}(z)=\frac{a_f^{\Lambda_c}}
{a_0^{\bar{\Lambda}_c}z}z^{1+\alpha_R(0)-2\alpha_N(0)}
(1-z)^{-\alpha_{\psi}(0)+\lambda}, \\
&&{\cal D}_{1dd}^{\Lambda_c}(z)=\frac{a_f^{\Lambda_c}}
{a_0^{\bar{\Lambda}_c}z}z^{2\alpha_R(0)-2\alpha_N(0)}
(1-z)^{-\alpha_{\psi}(0)+\lambda+2(1-\alpha_R(0))}, \nonumber\\
&&{\cal D}_{1ds}^{\Lambda_c}(z)=\frac{a_f^{\Lambda_c}}
{2a_0^{\bar{\Lambda}_c}z}z^{-2\alpha_N(0)+3\alpha_R(0)-\alpha_{\phi}(0)}
(1-z)^{-\alpha_{\psi}(0)+\lambda+
2(1-\alpha_R(0))+\alpha_R(0)-\alpha_{\phi}(0)}, \nonumber\\
&&{\cal D}_{SJ}^{\Lambda_c}(z)=\frac{a_f^{\Lambda_c}}
{a_0^{\bar{\Lambda}_c}z}z^{1-\alpha_{SJ}(0)}
(1-z)^{-\alpha_{\psi}(0)+\lambda+2(1-\alpha_R(0)}, \nonumber\\
&&{\cal D}_{0uu}^{\Lambda_c}(z)={\cal D}_{0dd}^{\Lambda_c}(z)=
\frac{a_0^{\bar{\Lambda}_c}}{z}(1-z)^{-\alpha_{\psi}(0)+\lambda+
4(1-\alpha_N(0))}, \nonumber\\ 
&&{\cal D}_{0ds}^{\Lambda_c}(z)=\frac{a_0^{\bar{\Lambda}_c}}{z}
(1-z)^{-\alpha_{\psi}(0)+\lambda+4(1-\alpha_N(0))+
\alpha_R(0)-\alpha_{\phi}(0)}. \nonumber
\end{eqnarray}

\begin{eqnarray}
&&{\cal D}_{d}^{\bar{\Lambda}_c}(z)={\cal D}_{u}^{\bar{\Lambda}_c}(z)=
\frac{a_0^{\bar{\Lambda}_c}}{z}(1-z)^{\alpha_R(0)-2\alpha_N(0)+
\alpha_R(0)-\alpha_{\psi}(0)+\lambda+2(1-\alpha_N(0))}, \nonumber\\
&&{\cal D}_{\bar{u}}^{\bar{\Lambda}_c}(z)={\cal D}_{u}^{\Lambda_c}(z), \nonumber\\
&&{\cal D}_{s}^{\bar{\Lambda}_c}(z)={\cal D}_{s}^{\Lambda_c}(z), \\
&&{\cal D}_{0ud}^{\bar{\Lambda}_c}(z)={\cal D}_{0dd}^{\bar{\Lambda}_c}(z)=
\frac{a_0^{\bar{\Lambda}_c}}{z}(1-z)^{\alpha_{R}(0)-2\alpha_{N}(0))+\lambda+
2(1-\alpha_N(0))+\alpha_R(0)-\alpha_{\psi}(0)}, \nonumber\\
&&{\cal D}_{0ds}^{\bar{\Lambda}_c}(z)=\frac{a_0^{\bar{\Lambda}_c}}{z}
(1-z)^{\alpha_{R}(0)-2\alpha_{N}(0))+\lambda+
2(1-\alpha_N(0))+\alpha_R(0)-\alpha_{\psi}(0)+
\alpha_R(0)-\alpha_{\phi}(0)}. \nonumber
\end{eqnarray}

\newpage

\begin{figure}[th]
\centerline{\epsfig{figure=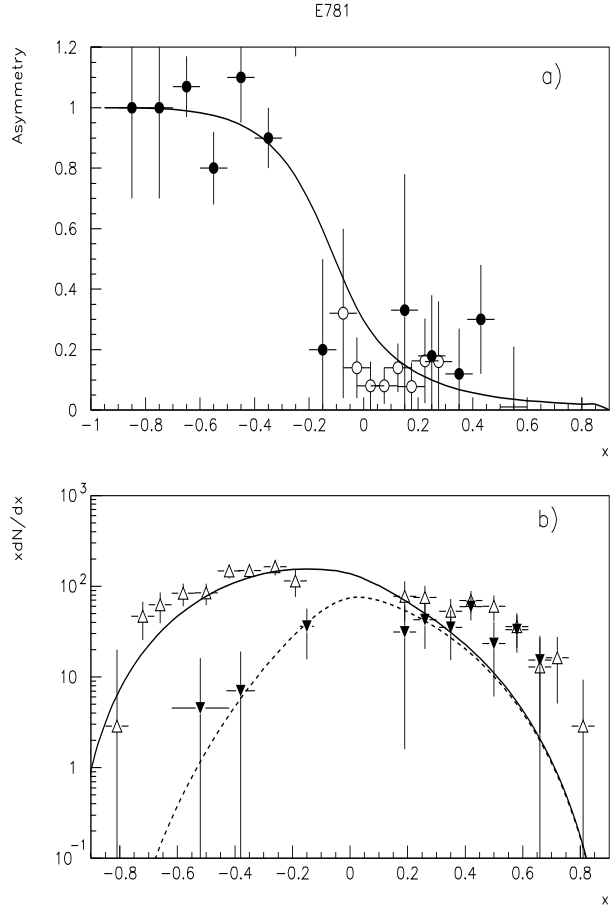,height=12cm,width=8cm}}
\caption{ a) Asymmetry between $\Lambda_c$ and 
$\bar{\Lambda}_c$ spectra 
obtained for $\pi-A$ ($x>0$) and for $p-A$ ($x<0$) collisions 
in the E781 expiment (black circles) \protect\cite{selex} and 
in the E791 experiment (empty circles) \protect\cite{e791},
the QGSM calculation with the string junction transfer (solid line); 
b) The distributions of $\Lambda_c$ (empty triangles) and 
$\bar{\Lambda}_c$ (black triangles) in E781 for these reactions 
and QGSM curves: $\Lambda_c$ (solid line) and $\bar{\Lambda}_c$ 
(dashed line).} 
\end{figure}

\begin{figure}[thb]
\centerline{\epsfig{figure=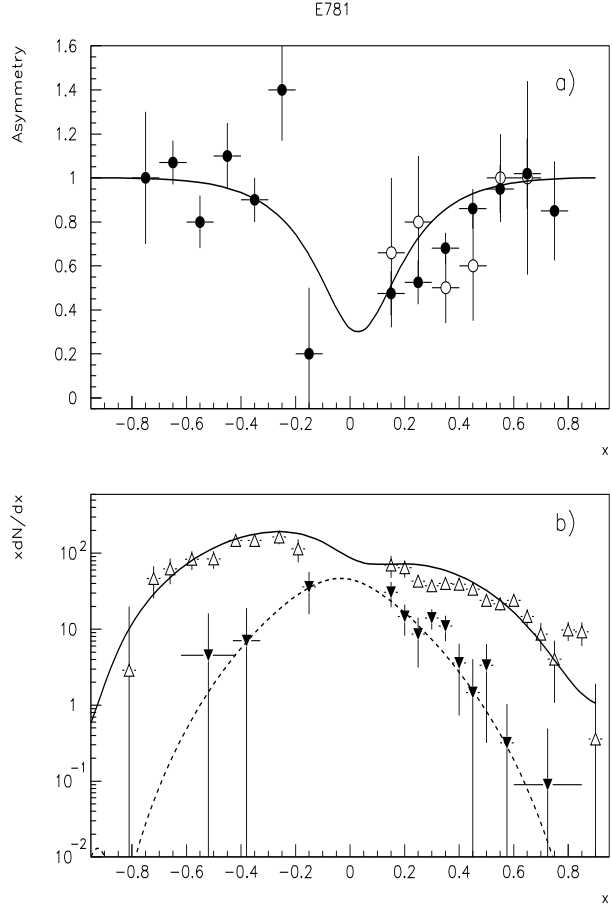,height=12cm,width=8cm}}
\caption{a) Asymmetry between $\Lambda_c$ and 
$\bar{\Lambda}_c$ spectra obtained for $\Sigma^--A$ ($x>0$) and 
for $p-A$ ($x<0$) collisions in the E781 experiment 
(black circles) \protect\cite{selex} and in the WA89 experiment (empty circles) 
\protect\cite{wa89}; the QGSM calculations (solid line);
b)The spectra of $\Lambda_c$ (empty triangles) and 
$\bar{\Lambda}_c$ (black triangles)
in E781 for these interactions and 
the corresponding QGSM curves: $\Lambda_c$ (solid line) and 
$\bar{\Lambda}_c$ (dashed line).} 
\end{figure}
\end{document}